\def\papertitle{Vocal Tract Area Estimation by Gradient Descent}
\def\paperauthorA{David Südholt}
\def\paperauthorB{Mateo Cámara}
\def\paperauthorC{Zhiyuan Xu}
\def\paperauthorD{Joshua D. Reiss}

\documentclass[twoside,a4paper]{article}
\usepackage{etoolbox}
\usepackage{dafx_23}
\usepackage{amsmath,amssymb,amsfonts,amsthm}
\usepackage{euscript}
\usepackage[T1]{fontenc}
\usepackage[utf8]{inputenc}
\usepackage{ifpdf}
\usepackage[english]{babel}
\usepackage{caption}
\usepackage{color}
\usepackage{booktabs}

\input glyphtounicode
\ninept

\newcounter{numauth}
\newcounter{listcnt}
\newcommand\authcnt[1]{\ifdefined#1 \stepcounter{numauth} \fi}

\newcommand\addauth[1]{
\ifdefined#1 
\stepcounter{listcnt}
\ifnum \value{listcnt}<\value{numauth}
\appto\authorslist{, #1}
\else
\appto\authorslist{~and~#1}
\fi
\fi}
\authcnt{\paperauthorB}
\authcnt{\paperauthorC}
\authcnt{\paperauthorD}
\authcnt{\paperauthorE}
\authcnt{\paperauthorF}
\authcnt{\paperauthorG}
\authcnt{\paperauthorH}
\authcnt{\paperauthorI}
\authcnt{\paperauthorJ}
\def\authorslist{\paperauthorA}
\addauth{\paperauthorB}
\addauth{\paperauthorC}
\addauth{\paperauthorD}
\addauth{\paperauthorE}
\addauth{\paperauthorF}
\addauth{\paperauthorG}
\addauth{\paperauthorH}
\addauth{\paperauthorI}
\addauth{\paperauthorJ}

\usepackage{times}

\newif\ifpdf
\ifx\pdfoutput\relax
\else
   \ifcase\pdfoutput
      \pdffalse
   \else
      \pdftrue
\fi

\ifpdf 
  \usepackage[pdftex,
    pdftitle={\papertitle},
    pdfauthor={\authorslist},
    pdfsubject={Proceedings of the 26th International Conference on Digital Audio Effects (DAFx23)},
    colorlinks=false, 
    bookmarksnumbered, 
    pdfstartview=XYZ 
  ]{hyperref}
  \usepackage[pdftex]{graphicx}
\else 
  \usepackage[dvips]{epsfig,graphicx}
  \usepackage[dvips,
    pdftitle={\papertitle},
    pdfauthor={\authorslist},
    pdfsubject={Proceedings of the 26th International Conference on Digital Audio Effects (DAFx23)},
    colorlinks=false, 
    bookmarksnumbered, 
    pdfstartview=XYZ 
  ]{hyperref}
\fi
\usepackage[hypcap=true]{caption}
\usepackage{subcaption}
\title{\papertitle}

\usepackage{tikz}
\usetikzlibrary{shapes.geometric, arrows, positioning}
\threeaffiliations{
\paperauthorA \,}
{Centre for Digital Music \\ Queen Mary University of London\\ London, UK\\
{\tt \href{mailto:d.sudholt@qmul.ac.uk}{d.sudholt@qmul.ac.uk}}
}
{\paperauthorB \,\sthanks{Work performed as part of an academic visit to the Centre for Digital Music, Queen Mary University of London}}
{Information Processing \& Telecomm. Center \\ Universidad Politécnica de Madrid \\ Madrid, Spain \\ {\tt \href{mailto:mateo.camara@upm.es}{mateo.camara@upm.es}}
}
{\paperauthorC,\, \paperauthorD}
{Centre for Digital Music \\ Queen Mary University of London\\ London, UK\\ {\tt 
\href{mailto:zhiyuan.xu@qmul.ac.uk}{zhiyuan.xu@qmul.ac.uk}
\href{mailto:joshua.reiss@qmul.ac.uk}{joshua.reiss@qmul.ac.uk}}
}

\begin{document}
\ifpdf 
  \DeclareGraphicsExtensions{.png,.jpg,.pdf}
\else  
  \DeclareGraphicsExtensions{.eps}
\fi


\maketitle

\begin{abstract}
Articulatory features can provide interpretable and flexible controls for the synthesis of human vocalizations by allowing the user to directly modify parameters like vocal strain or lip position. To make this manipulation through resynthesis possible, we need to estimate the features that result in a desired vocalization directly from audio recordings. In this work, we propose a white-box optimization technique for estimating glottal source parameters and vocal tract shapes from audio recordings of human vowels. The approach is based on inverse filtering and optimizing the frequency response of a wave\-guide model of the vocal tract with gradient descent, propagating error gradients through the mapping of articulatory features to the vocal tract area function. We apply this method to the task of matching the sound of the Pink Trombone, an interactive articulatory synthesizer, to a given vocalization. We find that our method accurately recovers control functions for audio generated by the Pink Trombone itself. We then compare our technique against evolutionary optimization algorithms and a neural network trained to predict control parameters from audio. A subjective evaluation finds that our approach outperforms these black-box optimization baselines on the task of reproducing human vocalizations.
\end{abstract}

\section{Introduction}
\label{sec:intro}

\begin{figure}
    \centering
    \includegraphics[width=0.83\columnwidth]{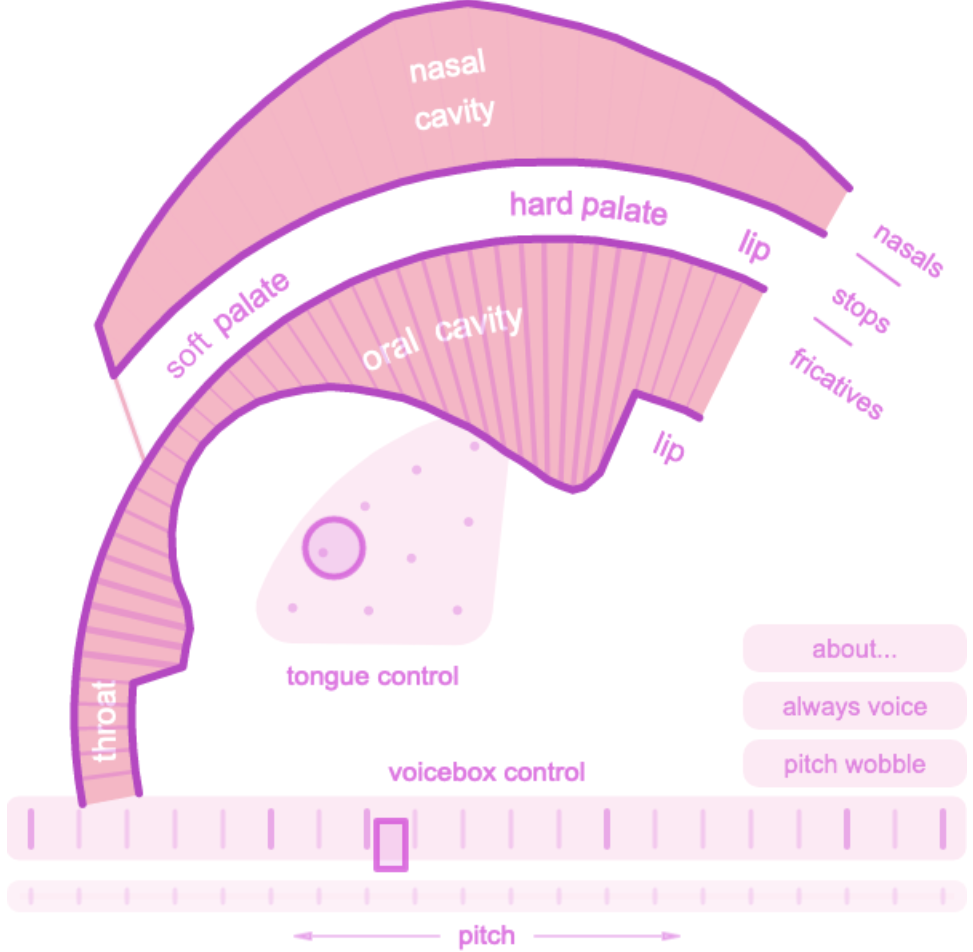}
    \caption{\emph{The user interface of the Pink Trombone articulatory synthesizer.}}
    \label{fig:ptui}
\end{figure}
Articulatory synthesis is a type of speech synthesis in which the position and movement of the human articulators, such as the jaw, lips or tongue, are used as control parameters. Because of their inherent interpretability, articulatory features lend themselves well towards fine-grained and flexible user control over the speech synthesizer \cite{birkholzModelingConsonantVowelCoarticulation2013}. Articulatory Synthesis is typically implemented as a physical model, which simulates the propagation of air pressure waves through the human vocal tract. A large number of such models have been developed over the years \cite{krogerComputerImplementedArticulatoryModels2022}. 

Obtaining the articulatory features that control the physical model is not a trivial problem. Area functions of the vocal tract can be directly measured with magnetic resonance imaging (MRI) \cite{storyVocalTractArea1996} or electromagnetic articulography (EMA) \cite{toutiosArticulatorySynthesisFrench2013}. However, these procedures are time-consuming, susceptible to noise and variations, and require access to specialized equipment. It is therefore desirable to recover the articulatory features directly from a given speech signal. In general, this task is known as Acoustic-to-Articulatory Inversion (AAI). Two main strands of research can be identified: one is data-driven AAI, which seeks to develop statistical methods based on parallel corpora of speech recordings and corresponding MRI or EMA measurements \cite{dangEstimationVocalTract2002,liuDeepRecurrentApproach2015}. The other takes an analysis-by-synthesis approach to AAI, in which numerical methods are developed to both obtain acoustic features from articulatory configurations, and to invert that mapping to perform AAI \cite{atalInversionArticulatoryAcoustic1978,sorokinEstimationStabilityAccuracy2000,richmondEstimatingArticulatoryParameters2001}.

In this work, we focus on the analysis-by-synthesis approach and consider the specific articulatory features that make up the control parameters of an articulatory synthesizer. The AAI task is then framed as obtaining control parameters such that the synthesizer reproduces a target recording. This allows a user to reproduce that vocalization with the articulatory synthesizer, and then modify parameters such as vocal tract size, pitch, vocal strain, or vowel placement.

Attempts to solve this problem of \emph{sound matching}, for articulatory synthesis or other types of synthesis, can generally be classified into \emph{black-box} and \emph{white-box} methods. 

Black-box methods do not rely on information about the structure of the synthesizer. A popular approach is to use derivative-free optimization techniques such as genetic algorithms \cite{riionheimoParameterEstimationPlucked2003, cooperSingingSynthesisEvolved2006, schleusingJointSourceFilterOptimization2013, gaoArticulatoryCopySynthesis2019, masudaQualityDiversitySynthesizer2021} or particle swarm optimization \cite{ismailVocalTractArea2008}. These methods are computationally expensive and can take many iterations to converge to a solution. Various deep neural network (DNN) architectures have also been proposed to predict control parameters that match a given sound \cite{gabrielliIntroducingDeepMachine2017, yee-kingAutomaticProgrammingVST2018, sahaLearningJointArticulatoryacoustic2020, shibataUnsupervisedAcoustictoArticulatoryInversion2021, martinezramirezDifferentiableSignalProcessing2021}. They require constructing high-quality datasets for training that cover the space of acoustic outputs.

White-box methods can improve the sound matching of specific synthesizers by incorporating knowledge of their internal structure. This can be done by reasoning about their underlying physical processes \cite{chatziioannouEstimationClarinetReed2012, wilkinsonLatentForceModels2017} or, more recently, making use of auto-differentiation and gradient descent techniques \cite{engelDDSPDifferentiableDigital2020, colonelDirectDesignBiquad2022, caspeDDX7DifferentiableFM2022, diazRigidBodySoundSynthesis2022}.

\begin{figure*}
    \centering
    \includegraphics[width=\textwidth]{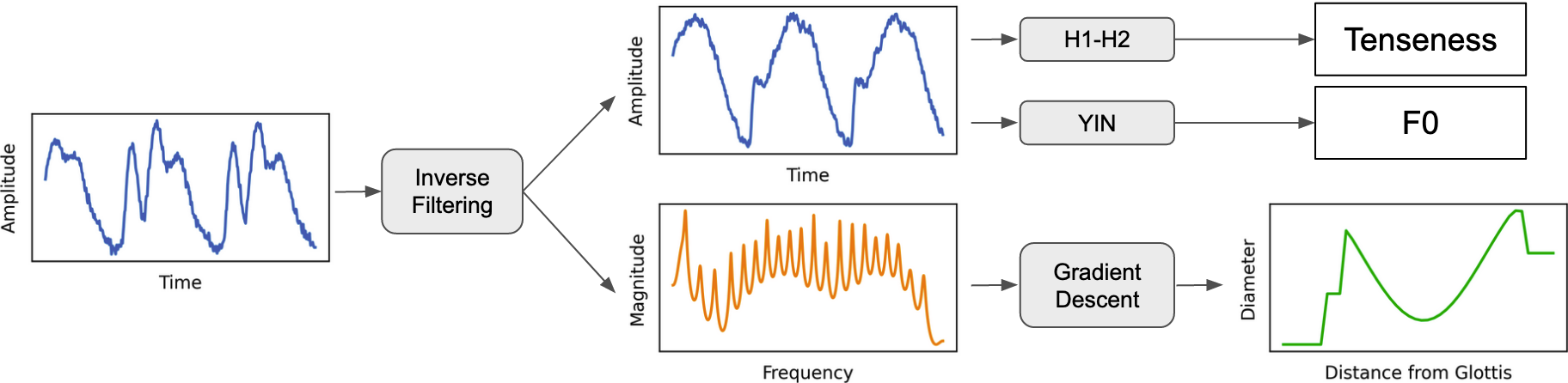}
    \caption{\emph{Illustration of the proposed sound matching method. Target audio is inverse filtered to obtain a source waveform and the transfer function of a filter. For resynthesis, the glottal control parameters $F_0$ and Tenseness are estimated from the source waveform. The vocal tract area function is optimized with gradient descent to match the filter's transfer function.}}
    \label{fig:overview}
\end{figure*}

In this work, we propose a gradient-based white-box optimization technique for sound matching vowel sounds with the articulatory synthesizer known as the Pink Trombone (PT)\footnote{\label{ptnote}\url{https://dood.al/pinktrombone}}. The PT is a web application that uses well-known models of the glottal source and the vocal tract to implement an intuitively controllable vocal synthesizer. Its user interface is depicted in Figure~\ref{fig:ptui}.

Our technique works as follows. First, we decompose a recording into a glottal source signal and an IIR filter with existing inverse filtering methods. We then obtain a vocal tract configuration by minimizing the difference between an analytical formulation of the tract's transfer function \cite{smythTransferFunctionPiecewiseCylindrical2021} and the IIR filter with gradient descent. A differentiable implementation of the mapping between control parameters and the vocal tract configuration allows propagation of the error gradient directly to the control parameters. Section~\ref{sec:soundmatching} describes the details of our approach.

We find that this approach can accurately recover the vocal tract area function on vowel sounds generated by the PT itself. A subjective listening test shows that without requiring any training procedures, the approach outperforms black-box baselines on the task of reproducing real human vocalization. The results of the objective and subjective evaluations are presented in section~\ref{sec:results}. Section~\ref{sec:conclusion} concludes the paper.

\section{Method}\label{sec:soundmatching}

The PT is based on the widely used source-filter model of speech production. The speech output $S(z) = G(z)V(z)L(z)$ is assumed to be the combination of three linear time-invariant (LTI) systems: the glottal flow $G$, the vocal tract $V$, and the lip radiation $L$. The lip radiation is approximated as a first-order differentiator $L(z) = 1 - z^{-1}$ and combined with $G$ to form a model of the \emph{glottal flow derivative} (GFD). Speech is then synthesized by generating a GFD signal (the source) and filtering it through the vocal tract $V$. 

In our sound matching approach, a target sound is first decomposed into the GFD source waveform and coefficients for an all-pole filter, using the inverse filtering technique proposed in \cite{perrotinSpectralGlottalFlow2019}. The control parameters for the PT glottal source are then obtained directly from the GFD waveform. We propose an objective function based on the magnitude response of the all-pole filter that allows estimating the control parameters of the vocal tract with gradient descent. The overall method is illustrated in Figure~\ref{fig:overview}. The source code is available online\footnote{\url{https://github.com/dsuedholt/vocal-tract-grad}}.

\subsection{Inverse Filtering}

To separate target audio into a GFD waveform and a vocal tract filter, we use the Iterative Adaptive Inverse Filtering method based on a Glottal Flow Model (GFM-IAIF) \cite{perrotinSpectralGlottalFlow2019}. 

IAIF methods in general obtain gross estimates of $G$, $V$ and $L$ with low-order LPC estimation, and then iteratively refine the estimates by inverse filtering the original audio with the current filter estimates, and then repeating the LPC estimation at higher orders.

GFM-IAIF makes stronger assumptions about the contribution of the glottis $G$, and uses the same GFD model as the PT synthesizer (compare section~\ref{sec:glottis}), making it a good choice for our sound matching task.

From GFM-IAIF, we obtain an estimate for the vocal tract filter $V$ in the form of $N+1$ coefficients $a_0,\ldots a_N$ for an all-pole IIR filter:

\begin{equation}
    V(z) = \frac{1}{\sum_{i=0}^{N}a_iz^{-i}}
\end{equation}
This also gives us an estimate of the GFD waveform by inverse filtering the original audio through $V$, i.e.\ applying an all-zero FIR filter with feed-forward coefficients $b_i=a_i$.

\subsection{Glottal Source Controls}\label{sec:glottis}

\begin{figure}
    \centering
    \includegraphics[width=\columnwidth]{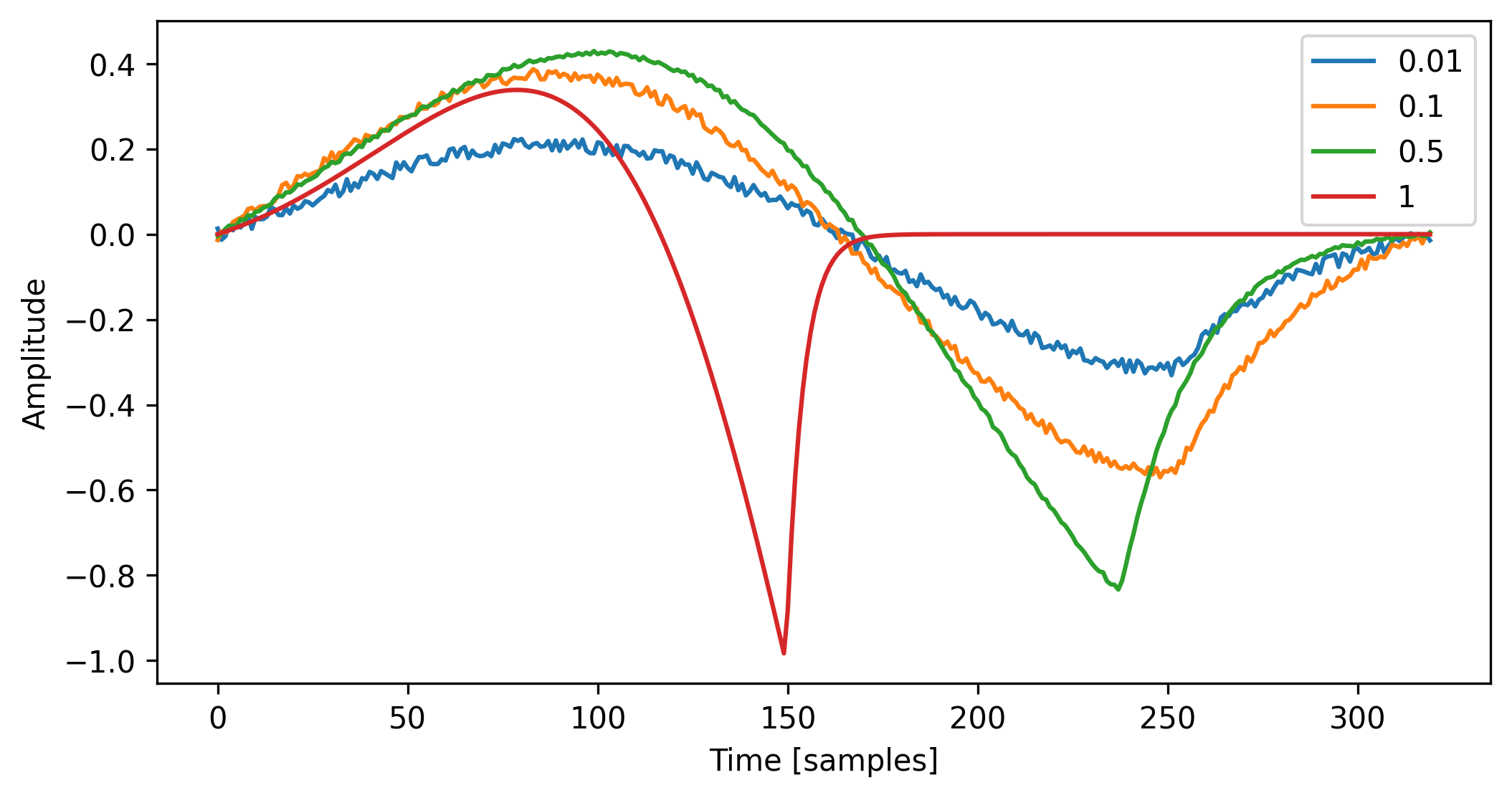}
    \caption{\emph{A single cycle of the glottal source waveform of the Pink Trombone, which combines the LF model with white noise, at varying values of the Tenseness parameter.}}
    \label{fig:lfmodel}
\end{figure}

The PT uses the popular Liljencrants-Fant (LF) model to generate the GFD waveform. Originally proposed with four parameters \cite{fantFourparameterModelGlottal1985}, the LF model is usually restated in terms of just a single parameter $R_d$, which is known to correlate well with the perception of vocal effort \cite{luGlottalSourceModeling2000}. 

$R_d$ can be obtained from the spectrum of the GFD. Specifically, \cite{fantLFmodelRevisitedTransformations1995} finds the following linear relationship between $R_d$ and $H_1-H_2$, the difference between the magnitudes of the first two harmonic peaks of the GFD spectrum (measured in dB):

\begin{equation}
    H_1-H_2 = -7.6 + 11.1R_d
\end{equation}
We estimate the fundamental frequency $F_0$ using the YIN algorithm \cite{DeCheveigne2002}, and measure the magnitudes of the GFD spectrum at the peaks closest to $F_0$ and $2\cdot F_0$ to calculate $H_1-H_2$ and thus $R_d$.

However, the PT does not use $R_d$ as a control parameter directly. Instead, it exposes a ``Tenseness'' parameter $T$, which relates to $R_d$ as $T = 1 - R_d/3$.

$T$ is clamped to values between 0 and 1, with higher values corresponding to higher perceived vocal effort. Additionally, the PT adds white noise with an amplitude proportional to $1 - \sqrt{T}$ to the GFD waveform, to give the voice a breathy quality at lower vocal efforts. Figure~\ref{fig:lfmodel} shows the glottal source at varying Tenseness values.

The estimated control parameters $F_0$ and Tenseness correspond to the horizontal and vertical axes in the PT's ``voicebox'' UI element, respectively (see Figure~\ref{fig:ptui}).

\subsection{Vocal Tract}

While the glottal source affects voice quality aspects like breathiness and perceived effort, the vocal tract is responsible for shaping the source into vowels and consonants.

In the PT, the vocal tract is treated as a sequence of $M+1$ cylindrical segments, with $M=43$. The shape of the vocal tract is then fully described by its \emph{area function}, i.e.\ the individual segment cross-sectional areas $A_0,\ldots, A_M$. Noting that $A = \pi(d/2)^2$, the area function may equivalently be described by the segment diameters $d_0,\ldots,d_M$. 

An additional, similar model of the nasal tract is coupled to the vocal tract at the soft palate. However, for the open vowel sounds that we are considering, the soft palate is closed and the coupling effect is negligible. In the PT implementation, the soft palate only opens when parts of the vocal tract are fully constricted, therefore here we focus only on the vocal tract itself.

\subsubsection{Control Model}

\begin{figure}
    \centering
    \includegraphics[width=\columnwidth]{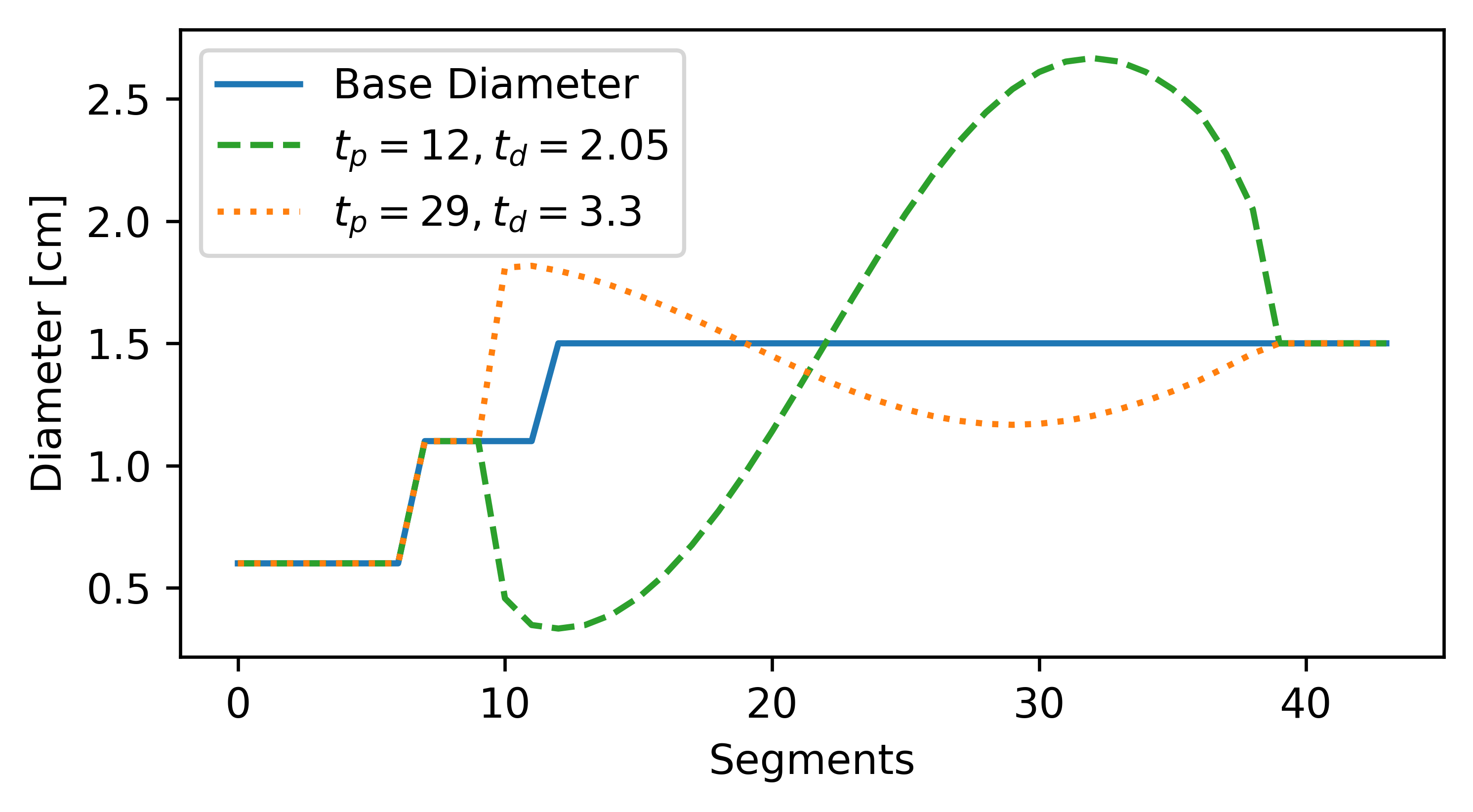}
    \caption{\emph{Example plots of the rest diameter, i.e.\ the result of applying the tongue model to the base diameter, at different tongue positions $t_p$ and tongue diameters $t_d$.}}
    \label{fig:tongue}
\end{figure}

Directly specifying each segment diameter individually does not make for an intuitive user experience and could easily result in very unrealistic, strongly discontinuous area functions. Instead, the PT implements a tiered control model over the vocal tract based on the model proposed in \cite{storyParametricModelVocal2005}. 

The control model consists of two tiers. The first tier is a tongue defined by a user-specified diameter $t_d$ and position $t_p$. The tongue shape is modeled as sinusoid shape and modifies a \emph{base diameter}, representing a neutral area function, into the \emph{rest diameter}. Figure~\ref{fig:tongue} illustrates this.

The second control tier are \emph{constrictions} that the user can apply to the rest diameter at any position along the vocal tract. Similarly to the tongue, constrictions are defined by an index, a diameter, and a model of how they affect the rest diameter. There are however two differences between the tongue and the constrictions: Firstly, constrictions are optional, while the tongue is always present. Secondly, constrictions can fully close the vocal tract, at which point noise is inserted to model plosives and fricatives. For this work, we consider only open area functions, meaning that we do not allow constrictions to reduce the diameter below a certain threshold.

\subsubsection{Estimating the Area Function}\label{sec:kldescription}

\begin{figure}
\centering
\includegraphics[width=\columnwidth]{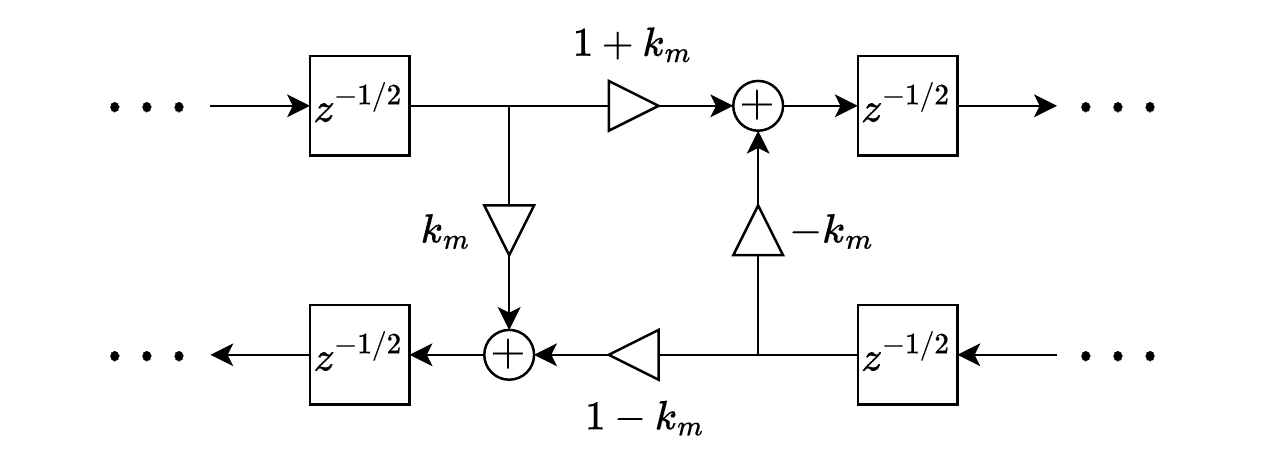}
\caption{\emph{Block diagram of a scattering junction in the Kelly-Lochbaum model, with scattering coefficient $k_m$.}}
\label{fig:klmodel}
\end{figure}

Propagation of the glottal source through the vocal tract is modeled by implementing each cylindrical segment as a bidirectional, half-sample delay. The half-sample delay is achieved by processing the signal at twice the audio sampling rate and adding up adjacent pairs of samples. At the $M$ inner junctions, the change in cross-sectional area leads to reflection and refraction, described by scattering coefficients calculated from the segment areas as

\begin{equation}\label{eq:reflcoeffs}
    k_m = \frac{A_m-A_{m-1}}{A_m+A_{m-1}}\,\text{ for }\,m=1,\ldots M.
\end{equation}
This is the well-known Kelly-Lochbaum (KL) model \cite{kellySpeechSynthesis1962}. An illustration of a scattering junction is shown in Figure~\ref{fig:klmodel}.

The length of the simulated vocal tract results from the number of segments and the sampling rate. Considering a speed of sound in warm air of $c \approx 350 \text{ m}/\text{s}$ and an audio sampling rate of $f_s = 48000\text{ Hz}$, implementing half-sample delays as unit delays processed at $2\cdot f_s$, $M + 1 = 44$ segments result in a vocal tract length of $44 \cdot 350 / (2\cdot48000) \approx 0.16\text{ m}$. This corresponds to the vocal tract of an average adult male \cite{storyParametricModelVocal2005}, giving the PT a male voice. The number of segments and the unit delays are fixed in the PT. The KL model can be implemented more flexibly through e.g.\ the use of fractional delays \cite{valimakiImprovingKellyLochbaumVocal1994}.

An analytical transfer function for the piecewise cylindrical model using unit delays was derived in \cite{smythTransferFunctionPiecewiseCylindrical2021}. The formulation can be straightforwardly adapted to half-sample delays by replacing every delay term $z^{-n}$ with $z^{-n/2}$, and then applying an additional factor of $1 + z^{-1}$ to account for the summing of adjacent samples. The transfer function $H_{\text{KL}}$ can then be stated as:

\begin{equation}\label{eq:kltf}
    H_{\text{KL}}(z) = \frac{(1 + z^{-1})z^{-(M+1)/2}\prod^M_{m=1}(1 + k_m)}{K_{1, 1} + K_{1, 2}R_L - R_0(K_{2, 1} + K_{2, 2}R_L)z^{-1}}
\end{equation}
$R_0$ and $R_L$ are the amount of reflection at the glottis and lips, respectively, and $K\in\mathbb{R}^{2\times2}$ is defined as follows:
\begin{equation}
    K = \begin{bmatrix}
        K_{1, 1} & K_{1, 2} \\
        K_{2, 1} & K_{2, 2}
    \end{bmatrix} = \prod_{m=1}^M\begin{bmatrix}
        1 & k_mz^{-1}\\
        k_m & z^{-1}
    \end{bmatrix}
\end{equation}
We now wish to find the tongue controls and constrictions such that $|H_{\text{KL}}|$ approximates $|V|$, the magnitude response of the vocal tract recovered by inverse filtering.

\begin{figure}
    \centering
    \includegraphics[width=0.6\columnwidth]{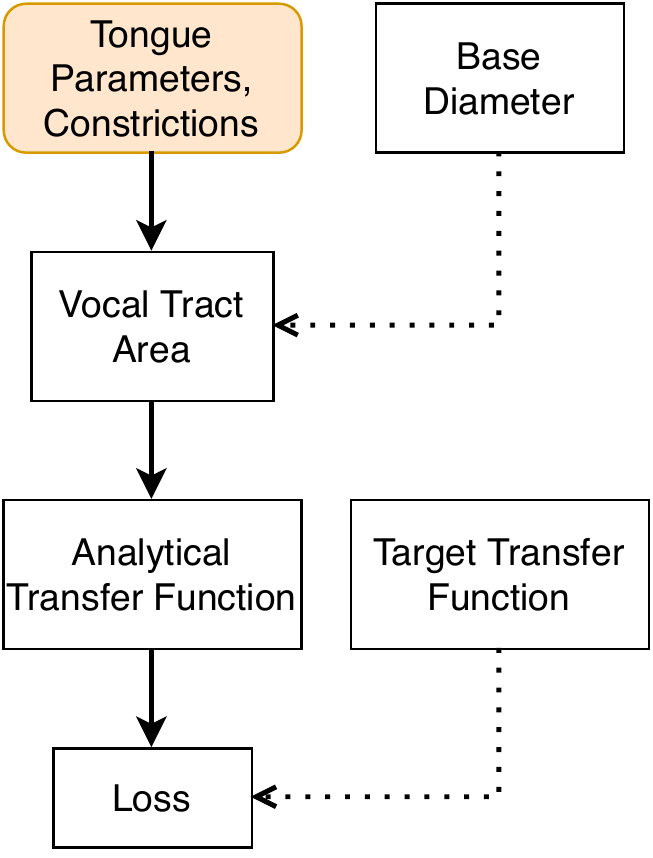}
    \caption{\emph{Schematic overview of the computation graph. In the forward pass, an area function is calculated from the control parameters. The corresponding transfer function is then computed and used to calculate the loss. Solid arrows denote that the operations are implemented to support auto-differentiation. This allows updating the estimate of control parameters (tongue and constrictions) using the gradient of the loss.}}
    \label{fig:graddescent}
\end{figure}

In an approach inspired by \cite{colonelDirectDesignBiquad2022}, we now consider the squared error between the log of the magnitude responses for a given angular frequency $0 \leq \omega < \pi$:

\begin{equation}
    E(\omega) = \left(\log_{10}|H_{\text{KL}}(e^{i\omega})| - \log_{10}|V(e^{i\omega})|\right)^2
\end{equation}
We can then define a loss function that measures how closely a given vocal tract area function matches the recovered vocal tract filter by evaluating the mean squared error over a set of $F$ linearly spaced frequencies:
\begin{equation}
    \mathcal{L} = \frac{1}{F}\sum_{f=0}^{F-1}E(\frac{f}{F}\pi)
\end{equation}
We can then find the set of controls that minimizes $\mathcal{L}$, meaning that the corresponding area function approximates $|V|$. A schematic overview of the computation graph is shown in Figure~\ref{fig:graddescent}.

\section{Experiments and Results}\label{sec:results}

We first evaluated the performance of our approach on recovering control parameters for sounds generated by the PT itself. These \emph{in-domain} sounds are guaranteed to be within the possible output space of the PT, and the ground truth parameters are known.

We then applied our approach to estimating control parameters for \emph{out-of-domain} sounds that were not generated by the PT itself. Ground truth parameters that provide an exact match are not known and likely do not exist due to limitations of the model, which makes evaluation challenging. We performed a listening test to compare the quality of our method to previously proposed, model-agnostic black-box sound matching approaches.

For all evaluations, parameter ranges were normalized to $[0, 1]$. Gradient descent was performed for $100$ steps, with a step size of $10^{-4}$ and a momentum of $0.9$. 

\subsection{Reconstructing PT-generated Audio}

\begin{table*}[!h]
\centering
\caption{\emph{MAE values for recovering control parameters when the target transfer function of the vocal tract (VT) is either given from the ground truth area function, or obtained by inverse filtering (IF). $t_p\in[12, 29]$ is the (continuous) position of the tongue along the vocal tract. $t_d\in[2.05, 3.5]$ is the tongue diameter.}\label{tab:indomain}}
\begin{tabular}{|l|cc|cc|cc|}
\hline
\# of Constrictions         & \multicolumn{2}{c|}{0} & \multicolumn{2}{c|}{1} & \multicolumn{2}{c|}{2} \\ \hline
VT Transfer Function        & Given      & IF        & Given      & IF        & Given      & IF        \\ \hline
$t_p$ {[}-{]}               & 0.19       & 1.42      & 1.21       & 1.93      & 1.74       & 2.15      \\
$t_d$ {[}cm{]}              & 0.02       & 0.26      & 0.12       & 0.28      & 0.19       & 0.32      \\
Total Diameter {[}cm{]}     & 0.01       & 0.23      & 0.07       & 0.24      & 0.11       & 0.24      \\
Frequency Response {[}dB{]} & 0.13       & 2.09      & 0.60       & 2.33      & 0.87       & 2.50      \\ \hline
\end{tabular}

\end{table*}
\subsubsection{Setup}

For the in-domain evaluation, we generated 3000 total sets of control parameters and attempted to recover the vocal tract area. For all examples, $F_0$ was uniformly sampled from $[80, 200]$, the tenseness from $[0, 1]$, the tongue position $t_p$ from $[12, 29]$ (measured in segments along the tract), and the tongue diameter $t_d$ from $[2.05, 3.5]$. The range of $F_0$ roughly covers the pitch range of adult male speech, while the other control parameter ranges cover the range of possible values defined by the PT interface.

The parameters were divided in three sets of 1000 examples each. The first set was taken as-is. A random constriction, with position sampled from $[0, 43]$ and diameter sampled from $[0.3, 2]$, was applied to the vocal tract in the second set. Two such independently sampled constrictions were applied in the third set. 

For each example, we performed the gradient descent optimization twice with different targets: First, with the target response $|V|$ taken directly from the ground truth frequency response (FR) of the original vocal tract. Since this FR is guaranteed to be within the domain of the KL vocal tract model, it should be able to be matched very closely.

Second, with the target response $|V|$ recovered by the GFM-IAIF method. This is no longer guaranteed to have an exactly matching vocal tract configuration, so higher deviation is expected. However, since GFM-IAIF and the PT are based on similar assumptions about the source-filter model, the obtained target responses match the ground truth closely enough to be useful in recovering the original control parameters.

\subsubsection{Results}
\begin{figure*}[!h]
    \centering
    \includegraphics[width=\textwidth]{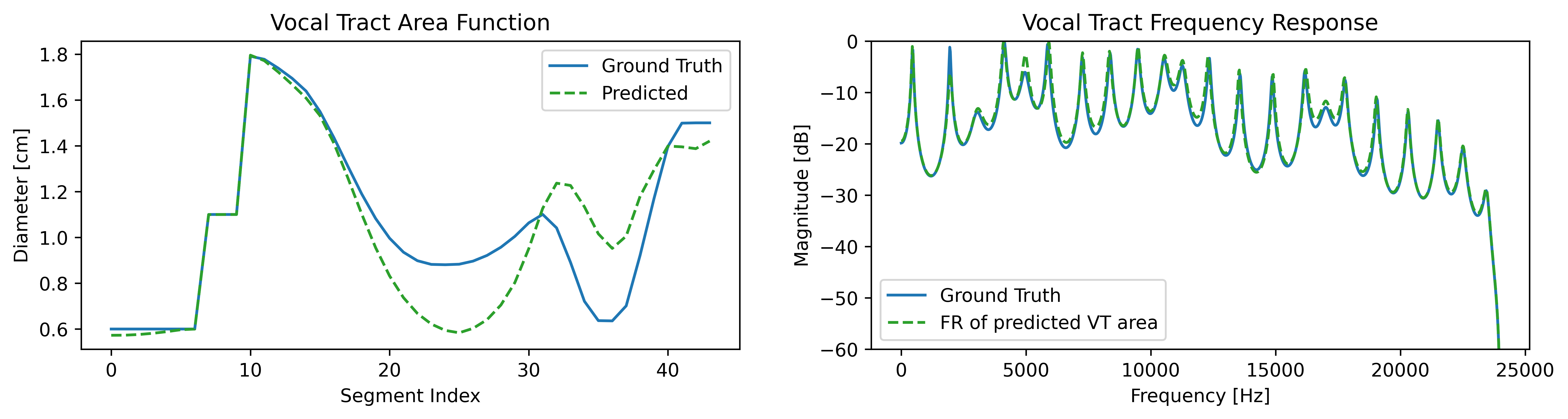}

     \caption{\emph{Visibly different area functions can have very similar frequency responses.} \label{fig:areavsfr}}
\end{figure*}

\begin{figure*}[!h]
    \centering
    \includegraphics[width=\textwidth]{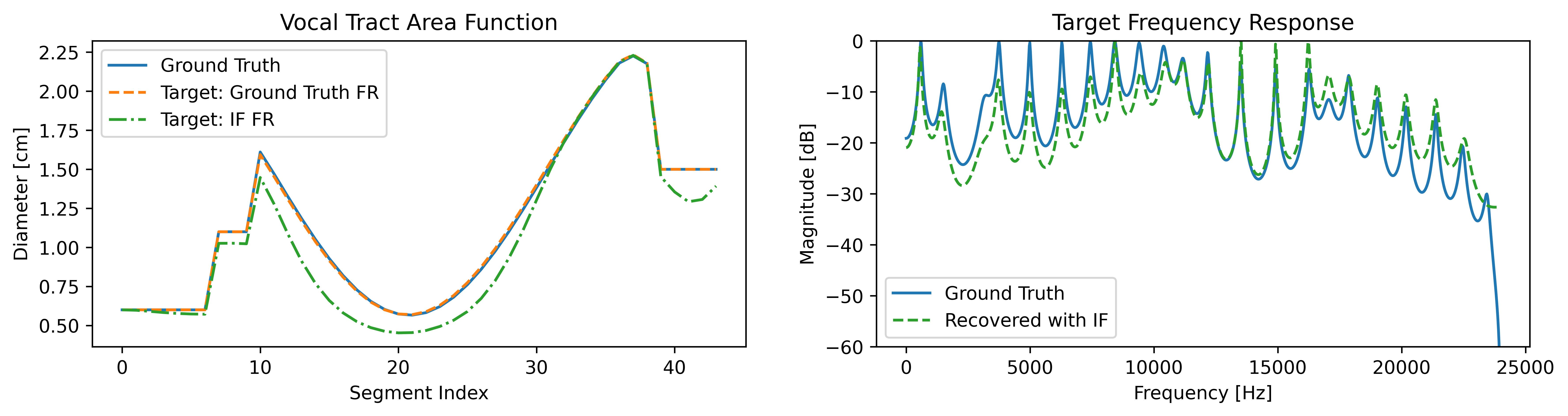}

     \caption{\emph{Area estimation results when either the frequency response (FR) of the true vocal tract or the result of inverse filtering (IF) are used as the target. The two different target frequency responses are shown on the right.} \label{fig:areaestimation}}
\end{figure*}

Table~\ref{tab:indomain} shows the mean absolute error (MAE) for the tongue parameters $t_p$ and $t_d$ for each condition. Additionally, the MAE values for the total area function (i.e.\ the diameter of each individual segment) and the recovered FR are given.

In the simple case of optimizing the true FR with no constrictions applied, the original vocal tract area could be recovered with very high accuracy, often to an exact match. Constrictions introduce more degrees of freedom and result in a less accurately recovered area function, although the FR was still matched very closely. Figure~\ref{fig:areavsfr} illustrates how visibly different area functions can have very similar frequency responses. This relates to the transfer function in equation~(\ref{eq:kltf}) not depending on the area directly, but rather on the resulting reflection coefficients in equation~(\ref{eq:reflcoeffs}). The locations of the area function's extrema, i.e.\ the segments at which the area changes from growing wider to growing more narrow or vice versa, therefore affect the transfer function more strongly than the specific value of a given area segment.

Since the FR obtained by GFM-IAIF might not be able to be matched exactly by the KL model, some constrictions might be used during the estimation even if there were none applied to the original vocal tract, leading to deviations from the true area function. An example of this is shown in Figure~\ref{fig:areaestimation}. The range of frequencies most affected by this depend on the choice of LPC estimation in GFM-IAIF; as noted in \cite{perrotinSpectralGlottalFlow2019}, modeling the glottal contribution as a $3^\text{rd}$ order filter is well-motivated by the LF model and gives balanced results in practice.

Due to the presence of this error introduced through inverse filtering, applying constrictions to the ground truth area function had a considerably less pronounced effect on the error metrics when the FR obtained by GFM-IAIF is used as the optimization target.

Inverse filtering also noticeably affected the estimation of the glottal source parameters. The MAE for the prediction of the tenseness $T\in[0, 1]$ was 0.013 when the original GFD waveform was used, but rose to 0.057 when the GFD waveform was recovered by inverse filtering. Even the accuracy of the YIN fundamental frequency estimator dropped slightly: the MAE for $F_0\in[80, 200]$ was 0.04 on the original GFD waveform, and 0.44 on the recovered GFD waveform.

Applying constrictions had no effect on the glottal source parameter estimation. Grouping the MAE values by the number of constrictions result in values deviating less than $0.5\%$ from the reported global MAE values for both $T$ and $F_0$. 

\subsection{Sound Matching Human Vocalizations}

\subsubsection{Black-Box Baselines}

To assess the out-of-domain performance, we performed a subjective evaluation comparing our gradient-based approach against three black-box optimization methods that have previously been used for the task of sound matching. 

\textbf{Genetic algorithms} \cite{riionheimoParameterEstimationPlucked2003, cooperSingingSynthesisEvolved2006, schleusingJointSourceFilterOptimization2013, gaoArticulatoryCopySynthesis2019, masudaQualityDiversitySynthesizer2021}  employ a population of candidate solutions, which evolve through generations by applying genetic operators such as selection, crossover, and mutation. The fittest individuals, evaluated through a fitness function, are more likely to reproduce and pass on their traits to offspring. 

\textbf{Particle Swarm Optimization} (PSO) \cite{ismailVocalTractArea2008} involves a group of candidate solutions, called particles, that move through the search space to find the global optimum. Each particle's position is updated based on its own best-known position, the best-known position within its neighborhood, and a random component, with the goal of balancing exploration and exploitation.

For both the genetic algorithm and PSO, scores for a given set of parameters were calculated as the mean squared error between the mel-spectrogram of the target audio, and the audio generated by the PT with the current parameters. 

\textbf{Neural parameter prediction} \cite{gabrielliIntroducingDeepMachine2017, yee-kingAutomaticProgrammingVST2018} uses a neural network to predict parameters from audio. We train a convolutional neural network (CNN) architecture with two convolutional layers separated by a max-pooling layer and followed by three fully connected layers on a dataset of 1,000,000 randomly sampled parameter sets and their corresponding mel-spectrograms.

While the in-domain evaluation focused on static vocal tract configurations, the speech samples used in the out-of-domain evaluation are time-varying. For all baselines and the gradient-based approach, this is handled by estimating the parameters on a frame-by-frame basis. To avoid sudden jumps in the area, the predictions of the baselines were smoothed over time by applying a Savitzky-Golay filter \cite{savitzkySmoothingDifferentiationData1964}. For our gradient approach, the estimation of each frame was initialized with the previous frame's prediction.

\subsubsection{Listening Test}

\begin{figure}
    \centering
    \includegraphics[width=0.95\columnwidth]{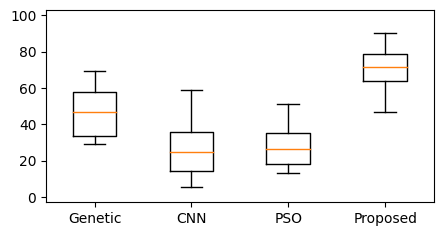}
    \caption{\emph{Boxplots showing the average rating across all stimuli of our gradient-based approach and black-box baselines.}\label{fig:listeningtest}}
    \label{fig:my_label}
\end{figure}

We reproduced 6 short recordings of human vocalizations with each method. The originals and the reproductions, and the individual ratings are available online.\footnote{\url{https://dsuedholt.github.io/vocal-tract-grad/}} The pitch, breathiness, and vowel shape of the recordings is time-varying. Each recording came from a different male speaker, since the PT's fixed vocal tract length limits its output to voices that are read as male (see section~\ref{sec:kldescription}). We set up an online multiple-stimulus test on the Go Listen platform \cite{barryGoListenEndtoEnd2021} asking participants to compare the four reproductions to the original recording and rate the reproduction on a scale of 0--100. We included an additional screening question in which we replaced one of the reproductions with the original recording to ensure participants had understood the instructions and were in a suitable listening environment.

22 participants took part in the listening test. Of those, 4 gave the original recording in the screening question a rating lower than 80, so their results were discarded. 

The results of the listening test are shown in Figure~\ref{fig:listeningtest}. Friedman's rank sum test indicates that the ratings differ significantly ($p < 0.001$), and post-hoc analysis using Wilcoxon's signed-rank test confirms that the reproductions obtained by our proposed approach are rated significantly ($p < 0.001$) higher than the three baselines, indicating that our method is well-suited for the sound matching task.

\section{Conclusion}\label{sec:conclusion}

We presented a white-box optimization technique for sound matching vowel sounds with the articulatory synthesizer. We obtained a vocal tract frequency response through inverse filtering and estimated corresponding articulatory control parameters with gradient descent optimization, propagating error gradients through the mapping of control parameters to the vocal tract area function.  We showed that our approach can accurately match frequency responses for audio generated by the synthesizer itself. Reproductions of time-varying human vocalizations generated with our approach outperformed black-box baselines in a subjective evaluation.

By showing that articulatory features can be estimated with a gradient-based method, our work lays the foundation for further research into end-to-end sound matching of articulatory synthesizers using neural networks, which require the propagation of gradients. Additionally, our method can be expanded to explore the sound matching of more complex synthesizers, including those with two- and three-dimensional vocal tract models and varying vocal tract lengths that are not limited to adult male voices.

\section{Acknowledgments}

This work was supported by UK Research and Innovation [grant number EP/S022694/1]. The authors would like to thank Benjamin Hayes, Yisu Zong, Christian Steinmetz and Marco Comunità for valuable feedback.

\bibliographystyle{ieeetr}
\bibliography{pinktrombonepaper} 

\end{document}